# Addressing the programming challenges of practical interferometric mesh based optical processors

KAVEH RAHBARDAR MOJAVER,[1,*] BOKUN ZHAO,[1] EDWARD LEUNG,[2]
S. MOHAMMAD REZA SAFAEE,[1] AND ODILE LIBOIRON-LADOUCEUR[1]

[1]*Electrical and Computer Engineering Department, McGill University, Montreal, Canada*
[2]*Department of Electrical and Computer Engineering, The University of British Columbia, Vancouver, Canada*
*\*hassan.rahbardarmojaver@mcgill.ca*

**Abstract:** We demonstrate a novel mesh of Mach-Zehnder interferometers (MZIs) for programmable optical processors. The proposed mesh, referred to as Bokun mesh, is an architecture that merges the attributes of the prior topologies Diamond and Clements. Similar to Diamond, Bokun provides diagonal paths passing through every individual MZI enabling direct phase monitoring. However, unlike Diamond and similar to Clements, Bokun maintains a minimum optical depth leading to better scalability. Providing the monitoring option, Bokun's programming is faster improving the total energy efficiency of the processor. The performance of Bokun mesh enabled by an optimal optical depth is also more resilient to the loss and fabrication imperfections compared to architectures with longer depth such as Reck and Diamond. Employing an efficient programming scheme, the proposed architecture improves energy efficiency by 83% maintaining the same computation accuracy for weight matrix changes at 2 kHz.

## 1. Introduction

Interferometric-based programmable optical processors are promising structures for fast and energy efficient computation in classic and quantum photonics [1–3]. These processors can efficiently perform analog vector-matrix multiplication from the inherent parallelism of optics [4]. Programmable optical processors can also do multiply-accumulate (MAC) operation in computing [5] and be used as quantum gates in quantum photonics [6]. With fast-growing computational demand in deep learning limiting its progress [7–8], energy efficient computational accelerators fabricated in silicon photonic (SiPh) technology is an excellent candidate to meet the computational demands of future machine learning and deep learning applications [9–10].

Ideally, the programmable optical processors should be fully reconfigurable after fabrication similar to what is offered by the electronics field-programmable gate arrays (FPGAs) [11–12]. However, there is a major difference between programmable optical processors and electronic FPGAs: the former is analog whereas the latter is digital. Performing analog computation includes several advantages mainly reducing the time and complexity of computation. The downside is that theses analog building blocks are more sensitive to the device parameters and bias condition. Fabrication variations and dynamic errors therefore translate into considerable computation error and inaccuracy in programmable optical processors [13–14]. Moreover, the processor performance drastically deteriorates with the dynamic errors including thermal and electrical crosstalk. One remedy to overcome the dynamic errors is to use *in-situ* programming [15–16]. The *in-situ* programming uses optimization techniques to find the optimum bias point of phase shifters for implementing a specific weight matrix. In *in-situ* programming, the processor is programmed in the presence of fabrication imperfections and dynamic errors, therefore, the phase shifters' bias can be tuned in a way to compensate for those errors. The downside is obvious; the time and energy required for programming. Each individual processor should go through a time/energy consuming optimization technique every time the weight matrix changes, and the optimization becomes drastically harder when scaling with matrix size.

The second option is *ex-situ* programming, *i.e.,* the phase shifters' bias corresponding to a specific weight matrix is externally calculated at first and then implemented on different similar chips. There are two main challenges with this approach. Firstly, hardware error correction schemes should be employed to compensate for the fabrication imperfections such as unbalanced splitting ratio [17]. Secondly, the processor should maintain an option to monitor the phase setting of phase shifters such that the phase setting can be fine-tuned to compensate for the dynamic errors. Waveguide taps or in-line transparent photodetectors may be considered to monitor the optical power and the state of numerous phase shifters on the processor [18]. However, these solutions often increase the insertion loss of the structure limiting the scalability. It would be favorable if monitoring the phase shift applied by a specific phase shifter within the optical paths is viable without modifying the bias of the other phase shifters.

In this work, for the first time, we propose an architecture, referred to as Bokun mesh, that provides direct phase monitoring while maintaining the minimum optical path depth. Bokun mesh is a topology arrangement that merges the attributes of two mesh topologies: Clements and Diamond [19, 20]. Providing the monitoring option, the programming would be faster improving the total energy efficiency of the processor. Bokun improves the total energy efficiency by 83% compared to the rectangular mesh for a $10 \times 10$ mesh with weight matrix changing at 2 kHz. The performance of Bokun mesh enabled by an optimal optical depth is also three times more resilient to the loss and fabrication imperfections compared to the architectures with longer depth such as Reck and Diamond for a $10 \times 10$ mesh used in a two-layered optical neural network for MNIST classification task [20, 21].

## 2. MZI-based optical processor architecture

Figure 1 (a) shows the building block of a Mach-Zehnder interferometer (MZI)-based programmable optical processors. This MZI is composed of two couplers (also referred as beam splitter/combiner) and two phase shifters. The linear transformation matrix of the $2 \times 2$ building block for a fixed state of polarization, 50:50 splitting ratio of couplers and assuming lossless optical propagation is:

$$\begin{bmatrix} O_{top} \\ O_{bottom} \end{bmatrix} = e^{j(\theta/2)} \begin{bmatrix} e^{j\varphi}\sin\left(\frac{\theta}{2}\right) & e^{j\varphi}\cos\left(\frac{\theta}{2}\right) \\ \cos\left(\frac{\theta}{2}\right) & -\sin\left(\frac{\theta}{2}\right) \end{bmatrix} \begin{bmatrix} I_{top} \\ I_{bottom} \end{bmatrix}, \quad (1)$$

where, $\theta$ is the internal phase shift changing the output optical intensity, and $\varphi$ is the external phase shift defining the output optical phase. $I_{top}$, $I_{bottom}$, $O_{top}$, and $O_{bottom}$, are the optical electric field distribution of a plane wave at the input and output ports [22].

An ideal $N \times N$ multiport reconfigurable MZI-based optical processor is a unitary optical component which consists of $n$ MZIs connected to each other based on a given mesh topology. The structure can be represented by an $N \times N$ unitary matrix implemented by the successive products of the unitary transformation matrices of its constituent MZIs. This process is done based on the location of the MZIs in the mesh. Unitary matrices preserve the inner product and norm of the transformed vectors, hence, they preserve the vectors length and angle. Unitary matrices have wide applications in machine learning, AI, and quantum computing. In deep learning, matrix multiplication is used to compute the activations of neurons in the neural network (NN). This involves multiplying the input data by weight matrices and adding biases to produce the output. When we use the optical processor to perform vector matrix multiplication of a NN, the $N$ input ports are employed to feed in the $d$ features of a sample

(where $N = d$). The $N$ output ports serve as the $c$-dimensional output vector to determine the single layer NN's predicted class of the sample. In this work we use two sample datasets. The multivariate Gaussian dataset already introduced in [20] and the MNIST dataset of handwritten digits [23]. We compare the results obtained from the two datasets to study the effect of dataset on hardware performance. The predicted class is determined by the index of an output representing the highest optical power. This is consistent with conventional NNs, where its predicted class is designated by the output neuron with the highest value [24].

The Reck mesh topology shown in Fig. 1 (b), theorized by Reck, *et al.* [21], consists of a triangular mesh of MZIs. It can be employed to implement an arbitrary $N \times N$ unitary matrix with $n$ MZIs connected to each other in a mesh with $N$ additional phase shifters at the input ports. In fact, an $N \times N$ Reck mesh is a universal $N \times N$ unitary transformation and it can be used to implement any $N \times N$ unitary transformation. The number of MZIs scale quadratically with respect to the size of the matrix as follows:

$$n = \frac{N(N-1)}{2}, \tag{2}$$

where, $N$ is the number of optical channels from the i$^{th}$ input to the j$^{th}$ output (i, j ∈ $N$) for a structure with the same number of inputs and outputs. The triangular architecture of Reck supports the sequential calibration of the mesh. In other words, the Reck architecture allows us to ensure every MZI is calibrated through a path of pre-calibrated MZIs.

The Diamond mesh shown in Fig. 1 (c), proposed by Shokraneh, *et al.* [20], employs $(N − 1) \times (N − 2)/2$ additional MZIs compared to the Reck and Clements meshes, for a total of MZIs given by:

$$n = (N − 1)^2. \tag{3}$$

While the additional number of MZI causes higher loss and greater susceptibility to phase uncertainty, it copes with loss imbalance due to increased symmetry of the mesh compared to the Reck mesh. Furthermore, it counteracts the effects of phase uncertainty by adding extra degrees of freedom in the created mesh through the additional MZIs. The unconnected output waveguides of the additional MZIs within the Diamond mesh allows for excluding the destructive portion of the interference from its outputs and optimally adjust the optical power levels at its outputs.

The Clements mesh depicted in Fig. 2 c, proposed by Clements, *et al.*, uses the same number of MZIs as the Reck mesh [19]. Although the total number of MZIs in Reck and Clements meshes is equal, Clements demonstrates shorter optical depth, therefore, less insertion loss. In Clements architecture, each input signal crosses its nearest neighbor at the first possible occasion leading to shorter optical depth. This is in contrast to the Reck mesh, where the bottom input signals must propagate for some distance before interacting with other signals. The mesh depth (number of consecutive MZIs in the longest path) for an $8 \times 8$ Clements topology is eight MZIs in contrast to 13 MZIs for the Reck. The downside of Clements topology is its complex calibration and programming. Unlike Reck and Diamond, the Clements structure is not triangular which makes its calibration challenging. In the next section, we show how triangular structures and presence of diagonal paths contribute to more precise calibration and programming.

In this work, we propose the novel Bokun mesh offering short optical depth while maintaining the triangular structure essential for accurate calibration and programming. Indeed, the Bokun mesh is a truncated diamond mesh with the middle optical I/Os used as the main optical path. It can also be considered as an extended version of Clements mesh with extra MZIs on the top and bottom of the structure to provide a triangular shape and offer a diagonal I/O path for each individual MZI. The number of MZIs in Bokun mesh is:

$$n = \frac{N(N+n/2-2)}{2}. \tag{4}$$

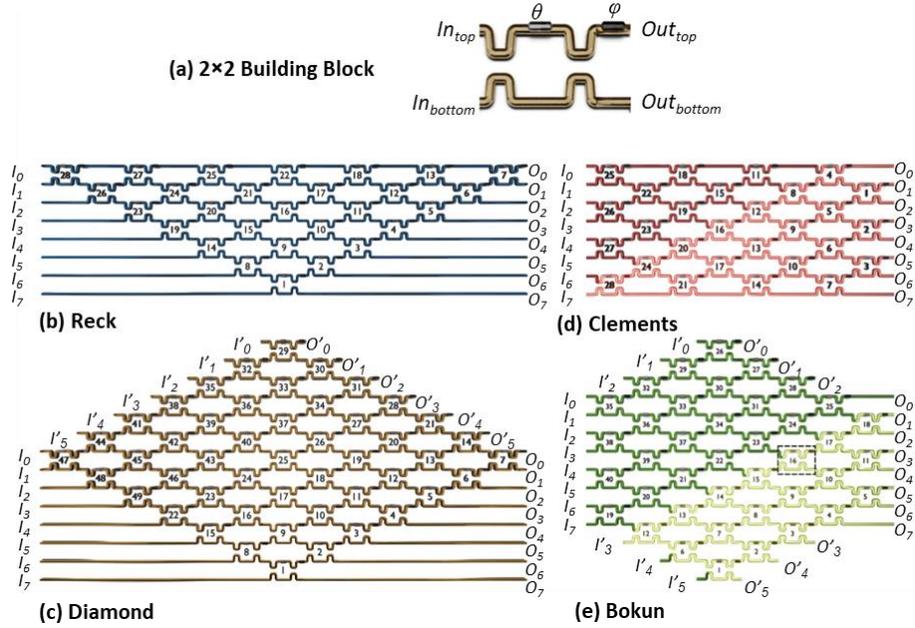

Fig. 1. (a) 2 × 2 building block of the processor. The 8 × 8 MZI-based optical processors as a (b) Reck, (c) Diamond, (d) Clements, and I Bokun (presented in this work) mesh. Reck is the primary interferometric architecture with a triangular architecture supporting sequential calibration scheme. The diamond is the symmetrical version of Reck with 21 extra MZIs. Clements reduces the mesh depth and provide more symmetry in terms of min/max number of MZIs in each path. Bokun mesh is an extended version of Clements with 12 extra MZIs and 12 auxiliary optical I/Os to mitigate the calibration/programming challenges. As discussed in section 3, an MZI is independently accessible if there is a way to light up the mesh so that one input of this MZI and all the subsequent MZIs towards an output remain dark (null). The illuminated section of the Bokun mesh highlights the monitoring of phase setting of MZI-16 by applying light only into $I'_3$ and detecting the light at $O_1$ while keeping all the rest of the inputs dark. In this condition, the top input of MZI-16 and secondary inputs of MZI-17 and MZI-18 remain null supporting independently monitoring of the MZI-16 phase setting. The figure also shows why this cannot be the case for the Clements mesh. In Clements, the light going through $I_7$ approaches both inputs of MZI-9 and the MZIs thereafter, so MZI-9 in Clements is not independently accessible. This is also the case if any other input is used. Among the four meshes, Bokun provides the shortest depth (equal with Clements), and the most balanced number of MZIs in paths (min=7 and max=8), while all its MZIs are independently accessible.

A unitary matrix can be implemented by the optical processor using its N MZIs by the inverse transformation matrices of the constituent MZIs, each being defined on an N-dimensional Hilbert space [21]. This allows for the matrix multiplication of an input vector representing a sample by injecting light at the input ports toward field interactions between the optical components in the mesh. The output vector is the result of the multiplication of the input vector by the unitary matrix.

## 3. Calibration and programming the optical processors

The first step in using programmable optical processors is the calibration. Through the calibration step, we find the relation between the bias voltage of phase shifters and the applied phase shift. In theory, similar phase shifters demonstrate identical phase/voltage relation. However in practice, the phase/voltage relation varies for similar phase shifters fabricated on the same chip due to the errors mainly coming from fabrication process variations [25]. Programmable optical processors are analog devices very sensitive to phase settings, therefore, one needs to calibrate each individual phase shifter prior to using the processor. In section 4, we will see how the phase error of a programmable optical processor degrades the classification accuracy of an optical neural network (ONN). In this section, we start discussing calibration of a single MZI as the 2 × 2 building block. Next, we extend the discussion to the system-level calibration of a larger mesh.

*3.1 Calibration of a single MZI*

Figure 2 (a) shows the calibration process of a single MZI. We start by calibrating the phase shifter $\theta$ and apply a continuous wave (CW) light into one input; the top input, *i.e.*, $In_{top}$ is chosen in this case. We then sweep the $\theta$ bias voltage and measure the optical power at a chosen output. We choose the bar state, therefore, top output ($Out_{top}$) in this measurement. The discussion would be similar for the cross state. According to eq. (1), one can find the optical electric field distributions from an input emitted to an output port. Focusing on the $In_{top}$ to $Out_{top}$ path, we write:

$$O_{top} = e^{j(\theta/2)} \left[ e^{j\varphi} \sin\left(\frac{\theta}{2}\right) I_{top} + e^{j\varphi} \cos\left(\frac{\theta}{2}\right) I_{bottom} \right], \tag{5}$$

Therefore, the transmission from the top input to the top output is:

$$Transmission = \left(\frac{O_{top}}{I_{top}}\right)^2 = \left(\sin\left(\frac{\theta}{2}\right) + \cos\left(\frac{\theta}{2}\right) \frac{I_{bottom}}{I_{top}}\right)^2, \tag{6}$$

For a precise calibration, we favor blocking the bottom input port ($In_{bottom}$) so no light passes through it. Under this condition, the $I_{bottom}$ in eq. (6) becomes zero and the transmission is proportional to $\sin^2\left(\frac{\theta}{2}\right)$ with a minimum and maximum at $\theta=0$ and $\theta=\pi$, respectively. We used Ansys Lumerical Interconnect to simulate the MZI transmission under various conditions. The black solid line in fig. 2 (b) presents the ideal case when no light goes to $In_{bottom}$. Knowing the corresponding bias voltage for $\theta=0$ and $\theta=\pi$, we calibrate the $\theta$ phase shifter.

In practice, if direct access to the MZI input is not viable, a small optical interference may be present at the $In_{bottom}$. Considering the optical powers of $P_{top}$ and $P_{bottom}$ at the top and bottom input ports with a relative phase shift of $\Delta_{in}$, we rewrite eq. (6) by plugging in $I_{top} = \sqrt{P_{top}}$ and $I_{bottom} = \sqrt{P_{bottom}} e^{j\Delta_{in}}$:

$$Transmission = \left(\sin\left(\frac{\theta}{2}\right) + \cos\left(\frac{\theta}{2}\right) \sqrt{\frac{P_{bottom}}{P_{top}}} e^{j\Delta_{in}}\right)^2, \tag{7}$$

Equation 7 formulates the effect of $P_{bottom}$ and $\Delta_{in}$ on $In_{top}$ to $Out_{top}$ transmission, hence, $\theta$ calibration error. Based on this equation, the calibration error increases with $P_{bottom}$ and maximized for $\Delta_{in}= 0, \pi$. Figure 3 (b) compares the $In_{top}$ to $Out_{top}$ transmission versus $\theta$, for different values of $P_{bottom}$ and $\Delta_{in}$. For $P_{bottom}$= -20 dBm and $\Delta_{in}$=0, the error in realizing the transmission minimum is 0.06 $\pi$. For $P_{bottom}$= -10 dBm, this error increases to 0.18$\pi$ and -0.18$\pi$, for $\Delta_{in}$=0 and $\pi$, respectively. In section 4, we will show that 0.18$\pi$ error on phase shifters drops the classification accuracy of a processor doing MNIST task by approximately 25%. The simulation results of Fig. 3(b) are in agreement with theory presented in eq. (7). We should note that the presence of a -10 dBm interfering signal at the bottom port is a realistic assumption, especially considering the fact that the previous stage of MZIs may not be calibrated [26].

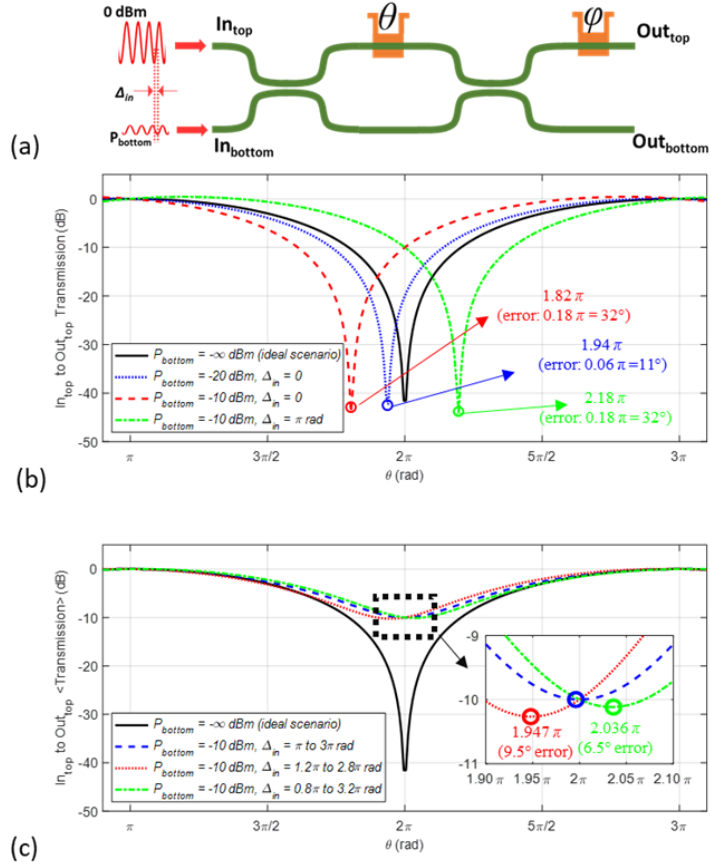

Fig. 2. (a) MZI calibration in the presence of an interfering signal at the bottom input. (b) Optical transmission versus $\theta$ for different cases of interfering signal at $In_{bottom}$. The calibration is ideal when $In_{bottom}$ is null (c) In the presence of a signal at $In_{bottom}$, changing $\Delta_{in}$ and averaging the optical transmission reduces the $\theta$ calibration error while increases the calibration complexity.

To remove the effect of interfering light at $I_{bottom}$ during the $\theta$ calibration, Bandyopadhyay, et al. discussed a useful method to average the transmission over $2\pi$ change of $\Delta_{in}$ [17]. Sweeping $\Delta_{in}$ can be done by using the external phase shifter of the previous MZI block. The average transmission is then equal to:

$$\langle Transmission \rangle = \frac{1}{2\pi}\int_0^{2\pi} \left( \sin\left(\frac{\theta}{2}\right) + \cos\left(\frac{\theta}{2}\right)\sqrt{\frac{P_{bottom}}{P_{top}}} e^{j\Delta_{in}} \right)^2 d\Delta_{in} = \sin^2\left(\frac{\theta}{2}\right), \quad (8)$$

As shown in fig. 2 (c), taking the average over a complete $2\pi$ period of $\Delta_{in}$, the transmission is minimized and maximized at $\theta=0$ and $\theta=\pi$, leading to error-free $\theta$ calibration. To ensure averaging over $2\pi$, the previous block controlling the $\Delta_{in}$ must be calibrated. If this is not the case, taking the average over a period slightly different than $2\pi$ would lead to a calibration error: $0.053\,\pi$ and $0.036\,\pi$ error for averaging over $2\pi - 0.4\pi$ and $2\pi + 0.4\pi$, respectively. The averaging technique contributes considerably to mitigating the calibration error even though it is not done exactly over a $2\pi$ shift of $\Delta_{in}$. The downside of the averaging technique is the increase in time and complexity of the calibration process. A typical thermo-optic phase shifter with $V_\pi \approx 2\,V$ [26] requires 400 measurement points for a $2\pi$ sweep with 0.01 V resolution.

Employing the averaging technique, assuming the same resolution, the number of measurement points goes up to 1600 for a two-dimensional sweep of $\theta$ and $\Delta_{in}$.

We now analyze the calibration error generated by the subsequent MZIs in an optical path towards its output. Figure 3 (a) shows MZI-1 (under calibration) connected to a photodetector through MZI-2. Based on eq. (7) and considering no light at $In_{bottom1}$ (ports labels are shown in fig. 3-a), the $In_{top1}$ to $Out_{top1}$ transmission is:

$$Transmission = \sin^2\left(\frac{\theta_1}{2}\right)\left(\sin\left(\frac{\theta_2}{2}\right) + \cos\left(\frac{\theta_2}{2}\right)\sqrt{\frac{P_{bottom_2}}{P_{top_2}}}e^{j\Delta_{in-2}}\right)^2, \qquad (9)$$

in which, $\theta_1$ and $\theta_2$ are the internal phase shift of MZI-1 and MZI-2, $P_{top2}$ and $P_{bottom2}$ are the optical power at the top and bottom input ports of second MZI with relative phase shift of $\Delta_{in-2}$. Ideally, we prefer the transmission to be proportional to $\sin^2\left(\frac{\theta_1}{2}\right)$, resulting in the minimum and maximum transmission at $\theta_1 = 0$ and $\theta_1 = \pi$, respectively. However, $P_{top_2}$ and $\Delta_{in-2}$ from the second part of eq. (9) are functions of $\theta_1$ leading to the calibration error.

To mitigate this error, the first solution is to set $\theta_2 = \pi$ (MZI-2 is in the bar state) making the cosine term in eq. (9) equal to zero. Thus, light at $In_{top2}$ directly goes to $Out_{top2}$ without interfering with $In_{bottom2}$. This scenario is shown in Fig. 4 (b) by the red dashed curve. The challenge is that any variation on $\theta_2$ translates into an error in the $\theta_1$ calibration. The second solution is to block the bottom input port of MZI ($P_{bottom_2} = 0$). According to eq. (9) the transmission would be $\sin^2\left(\frac{\theta_1}{2}\right)\sin^2\left(\frac{\theta_2}{2}\right)$, proportional to $\sin^2\left(\frac{\theta_1}{2}\right)$ for a constant value of $\theta_2$. Note that in this case, there is no need to precisely set $\theta_2$ to $\pi$ or 0. As long as $\theta_2$ is constant, MZI-2 acts as a constant optical attenuation between $In_{top1}$ to $Out_{top2}$ by tapping out a portion of signal to the $Out_{bottom2}$. The blue dotted curve in fig. 3 (b) presents this scenario.

If $P_{bottom_2} \neq 0$ and $\theta_2 \neq \pi$, MZI-2 adds error to the calibration of MZI-1. As shown by the dash-dotted green curve in fig. 4(b), for $P_{bottom_2} = -10\ dBm$, a 0.1 $\pi$ error in $\theta_2$ translates to a 0.02 $\pi$ error in $\theta_1$ calibration. Based on this discussion, to perform a precise calibration of an MZI, the second input port of all the MZIs in the following stages should be kept null.

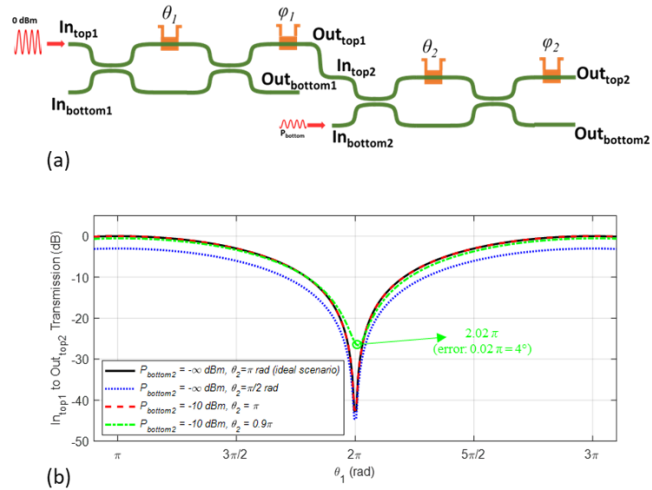

Fig. 3. (a) MZI calibration in the presence of a secondary MZI at its output. (b) Optical transmission versus $\theta_1$ for different conditions of interfering signal at $In_{bottom2}$. The calibration is ideal when $In_{bottom2}$ is null or $\theta_2 = \pi$

The external phase shifter $\varphi$ sets the output phase of the light coming out of the MZI, therefore, its calibration requires measuring the phase of signal. This can be done through employing multiple transverse electric (TE) modes or coherent detection of light [27]. For the phase shifter $\varphi$ calibration, the effect of interfering signal at the input of the MZI under calibration and all the subsequent MZIs is similar to what we discussed for the phase shifter $\theta$. In general, having one input of the MZI and all the following MZIs null contributes to precise calibration of both $\theta$ and $\varphi$ phase shifters. This is the principal advantage of diagonal optical paths in interferometric meshes as will be discussed in the next session.

*3.2 Calibrating a mesh of interferometers*

To calibrate a mesh of MZIs as shown in fig. 1, we need to calibrate each individual MZI. As discussed in the previous section, while calibrating an MZI in a mesh, we ideally need one input of the MZI null as well as one input of all MZIs null an the consecutive stages. This is viable for all MZIs on the Diamond and Bokun architectures thanks to the diagonal paths going through every MZIs in these meshes; however, Reck and Clements cannot provide this option.

To calibrate the Bokun mesh, we start with MZI-1 applying a light to $I'_5$ and connecting $O'_5$ to a detector. When no light goes through $I_7$, $I'_3$, and $I'_4$, we ensure the top input of MZI-1 is null. Next, we calibrate MZI-2 by applying a light to $I'_5$, setting MZI-1 in the cross state and connecting a detector to $O'_4$. Keeping $I_2$, $I_3$, $I_4$, $I_5$, $I_6$, $I_7$, and $I'_4$ dark, we ensure the top input of MZI-2 is null. We continue calibrating all MZIs in a similar manner with the sequence identified by MZI numbers in Fig. 2. The diamond shape architecture of the Bokun mesh provides a diagonal path (southwest to northeast or northwest to southeast) from an input to an output going through each individual MZIs. Taking this diagonal path while keeping all the other inputs dark ensures one input of the MZI and all the consecutive MZIs towards the output is null. The calibration process for MZI-16 as one of the middle MZIs is illustrated in Fig. 2. We illuminate $I'_3$, setting MZIs 12,13,14,15,17,18 all in their cross state. Note that due to the diagonal path, MZI-16 and the two consecutive MZIs (MZI-17 and MZI-18) have one null input. Therefore, the diamond shape architecture of the Bokun mesh provides the option for calibrating every MZI independently without phase error accumulation. The calibration of the Diamond mesh follows an almost similar procedure.

The MZIs in the Reck mesh shown in Fig. 2 can be calibrated in the sequence as enumerated. Due to the triangular topology of the Reck mesh, during the calibration procedure of an MZI one input can always be set as null. For example, in calibrating MZI-22, by applying an input to $I_4$, setting MZI-19, MZI-20 and MZI-21 in their cross state, and keeping $I_0$, $I_1$, $I_2$, and $I_3$ dark, we ensure the top port of MZI-22 is null. This helps eliminate the input error in MZI calibration. However, in Reck mesh, we cannot necessarily ensure one input of consecutive MZIs towards the output is null. In the example of calibrating MZI-22, whether we choose the path towards output through MZI-18, MZI-13, and MZI-7 or any other path towards output I/Os, all MZIs may have interfering signals at their secondary output generated from $I_4$. Although the triangular structure of Reck guarantees MZI-18, MZI-13, and MZI-7 are already calibrated during the calibration of MZI-22, any error on their phase setting translates to the calibration error of MZI-22. As discussed in the previous section, the error generated by the interfering optical light at the input of MZI under calibration is more severe than the error caused by the consecutive MZIs. Therefore, Reck calibration is fairly robust against phase errors. The downside of the Reck is the long and unbalanced optical depth as noted in section 2.

The calibration process of Clements is more elaborate. Clements architecture is designed in a rectangular shape so that each input signal crosses its nearest neighbor at the first possible occasion leading to minimum optical depth of the mesh. However, the short optical depth is at the price of a more involved and in some cases inaccurate calibration. Calibration of Clements is discussed in detail in [17]. The calibration starts from the last stage providing direct access to the outputs and continues towards the inputs with the sequence as enumerated in fig. 2. We start from MZI-1 and we choose its top input to top output path for calibration. While calibrating MZI-1, the structure does not provide any option to shine light at the top input of this MZI while keeping the bottom input dark. The light from the mesh input reaches MZI-1 through several non-calibrated MZIs. Therefore, the technique of averaging the input phase over 2π discussed in the previous section is used in calibrating MZI-1 to mitigate the effect of interfering light at the bottom input. Once the last stage MZIs are calibrated, the calibration process continues to the preceding stages of these mesh topologies.

*3.2 Programming and monitoring the state of phase shifters*

The programming is the processes of setting all the MZIs bias towards achieving a desired weight matrix. Unlike calibration which is done only one time after the fabrication of the programmable optical processor, the programming is done every time the weight matrix changes. Therefore, the time and energy consumption in the programming should not be ignored. In in-situ programming, the MZIs bias is set through an optimization technique such as gradient descent [15, 16]. In ex-situ programming, the bias points required for a specific weight matrix are externally calculated and implemented on different similar chips. In both techniques, every time the weight matrix changes, the processor should be reprogrammed. Therefore, the programming time should be less than one over the maximum frequency of the weight matrix change. This limits the application of the optical processor to stationary or low frequency variation weight matrix tasks. Once the weight matrix is applied to the phase shifters, dynamic errors caused by thermal crosstalk between the phase shifters, degrade the accuracy of the processor. The MZIs' bias must be readjusted to compensate for the thermal crosstalk generated by the adjacent phase shifters. The programming would be fast and more accurate if the processor included waveguide taps or in-line transparent photodetectors to monitor the phase setting of each MZI and provide a closed loop system for setting the phase shifters bias. However, waveguide taps and in-line photodetectors often increase the insertion loss of the structure and adds to the complexity of the system [11, 18]. Therefore, it would be a great advantage if the architecture can inherently provide an option for monitoring the phase shift applied by a specific phase shifter, through the main optical I/Os and without changing the other phase shifters' bias. The diagonal architecture of Bokun mesh provides this option.

Let us get back to the example of MZI-16 in the Bokun mesh presented in fig. 1. We showed earlier that by applying a light to $I'_3$, keeping all the other inputs dark, MZI-16 and all the consecutive MZIs towards $O_1$ have one dark optical input. We schematically demonstrate this situation in fig. 1, by lighting up the illustrated structure through applying light into $I'_3$. As discussed in section 3.1, keeping the bias of all MZIs except MZI-16 unchanged, the $I'_3$ to $O_1$ transmission is proportional to $\sin^2\left(\frac{\theta_{MZI-16}}{2}\right)$. When a specific weight matrix is implemented on the Diamond mesh, we can monitor $\theta_{MZI-16}$ without changing the bias of other MZIs. Because of this feature, we call MZI-16 independently accessible. This feature is very useful when we need to monitor the phase shift applied by a specific phase shifter after implementing the weight matrix. Since we do not need to change the state of other MZIs, we are able to monitor the state of MZI-16 in the presence of thermal crosstalk generated by the other MZIs. This is indeed not viable if we do not have a diagonal path between an input to an output going through MZI-16. Due to the diamond shape of Bokun, all MZIs in this structure are independently accessible.

In Clements, only the MZIs on the two diagonals are independently accessible. As an example, fig. 1 demonstrates why MZI-9 as an off-diagonal MZI is not independently accessible. If we shine light into $I_7$ and monitor $O_3$, both inputs of MZI-9 and the following MZIs towards the output are illuminated. After implementing a weight matrix on Clements, if we need to monitor the exact phase shift applied by the MZI-9, we need to change the configuration of MZI-12 and set them to cross state to null the upper input of MZI-9. We also must set MZI-5 and MZI-2 in the bar state so that the $I_7$ to $O_3$ transmission path becomes equal to a sinusoidal squared function of $\theta_{MZI-9}$. Similarly, in Reck, only the MZIs on the outer diagonal are independently accessible. Diamond, however, similar to Bokun has all its MZIs independently accessible.

Table 1 compares the characteristics of the four meshes shown in fig. 1. Diamond and Bokun support easier and more accurate programming having all their MZIs independently accessible. The main improvement in the Bokun compared to Diamond is the shorter depth. Indeed, Bokun is a truncated Diamond, in which we use the center optical I/Os as the main optical path. We also remove the MZIs on the two sides to minimize the optical depth. Bokun mesh can also be seen as a complementary Clements with extra MZIs on top and bottom. These MZIs are essential for providing diagonal path and increasing the number of independently accessible MZIs. Another advantage provided by Bokun is the balanced number of MZI among different optical paths (minimum of seven and maximum of eight in an $8 \times 8$ structure). The balanced number of MZIs, hence, balanced insertion loss is an important feature of a mesh especially in quantum applications [19]. The downside of Bokun, similar to Diamond, is the larger number of MZIs and optical I/Os. It should be noted that although larger number of MZIs increase the footprint of the mesh, the extra MZIs in Bokun (unlike Diamond) are not in the main optical path, keeping the mesh depth minimized.

Table 1. Architecture Characteristics of Different $8 \times 8$ Meshes

| Mesh | Total number of MZIs | Independently accessible MZIs | Mesh depth | Min↔Max MZIs per path |
|---|---|---|---|---|
| Reck [21] | 28 | 13 (46%) | 13 | 1↔13 |
| Diamond [20] | 49 | 49 (100%) | 13 | 1↔13 |
| Clements [19] | 28 | 14 (50%) | 8 | 4↔8 |
| Bokun (this work) | 40 | 40 (100%) | 8 | 7↔8 |

## 4. Performance of meshes in optical neural networks

To compare the performance of optical processors based on Reck, Clements, Diamond, and Bokun mesh, we simulate their performance when used as optical neural networks. A conventional single layer digital neural network is depicted in Fig. 4. The inputs (IN) are the features composing of a single sample of the dataset fed into the neural network through nodes **X**. The vector **X** is then multiplied by the weight matrix **W** before being sent through the nonlinear activation function $f()$, yielding a final vector $\hat{Y}$. As such, the equation for a single layer NN is:

$$\hat{Y} = f(Z) = f(W.X) \tag{10}$$

Vector $\hat{Y}$ is processed for the predicted class of the sample by finding its maximum argument. If the network is undergoing backpropagation [28], this $\hat{Y}$ can then be compared to the ground truth vector **Y**, which for classification purposes is generally a one-hot encoded vector [29]. The number of features is assumed to be equal to the number of classes. The comparison between the two vectors $\hat{Y}$ and $Y$ results in a loss function, $L$, such as a mean square error [30] or a categorical cross entropy [31]. Once this loss function value is calculated, the gradient is calculated with respect to the weight matrix. Subsequently, gradient descent is done on the network to optimize the weight matrix. The resulting weight matrix **W** can then be built using the optical processors described in the previous section.

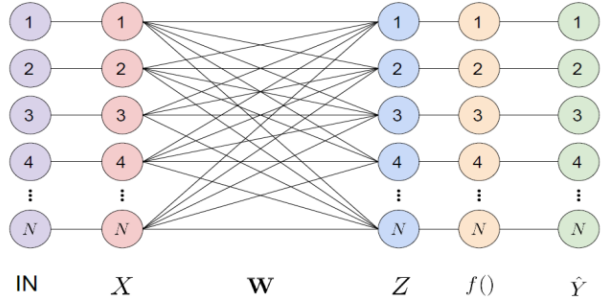

Fig. 4. Example of an N × N single layer neural network, taking in N features and returning N possible classes.

We use two datasets in this work to ensure our comparison is general enough and to ensure that one mesh is not favored due to specific characteristics of the used dataset. The first dataset used is the linearly separable Gaussian dataset presented in [20]. This dataset allows a conventional single layer NN to achieve a classification accuracy of 100% [32]. The second dataset is MNIST [23].

Neuroptica, written by the authors of [33], is a simulation platform for MZI-based ONNs written in Python. It provides a wide range of abstraction levels for training and simulating the ONNs. The lowest-level functionality is implemented allowing for the manipulation of the arrangement and properties of the phase shifters and the couplers of the MZIs, while the highest-level features provide a Keras-like application programming interface (API). This library allows for the training of MZI-based ONNs through backpropagation [15, 21]. It should be noted that the backpropagation algorithm used for ONNs must backpropagate all the way back to the phases of the MZIs rather than simply to the matrix weights. As such, the backpropagation algorithm uses the adjoint electric field method [22] to allow in situ optimization of the unitary transformation matrix of the mesh. The library includes both the Reck and the Clements meshes. However, it does not include the Diamond and Bokun mesh which were added allowing the training and simulation of the type of meshes studied in this work. The Diamond and Bokun meshes have been added in a cloned repository, Neuroptica: Towards a Practical Implementation of Photonic Neural Networks repository [34].

We test the performance of the meshes presented in Fig. 2, in the presence of two main sources of error, i.e, phase uncertainty and optical loss. Phase uncertainty in the phase shifters of each reconfigurable MZI is of great importance in the experimental programming of the MZI-based optical processors. The phase uncertainty of a phase shifter impacts the optical power splitting ratio at the outputs of the corresponding MZI. Phase error in the phase shifter corrupts the relative phase at its output ports. As a result, phase uncertainties degrade the classification accuracy of the implemented ONN. In this work, the phase uncertainty is represented by a normally distributed random variable ($N(0, \sigma)$). The phase uncertainty is affected by multiple factors, including thermal crosstalk between phase shifters, signal noise of the bias voltages applied to the phase shifters, and waveguide dimension variations. Thus, the phase shift in a phase shifter is mathematically defined as

$$\Theta_{actual} = \Theta_{optim} + N(0, \sigma_\Theta^2) \qquad (11)$$

where, $\Theta_{optim}$ is the optimal $\theta$ or $\varphi$ phase and $\sigma_\Theta$ is the standard deviation of either phase. This phase uncertainty is recalculated after each matrix multiplication to mimic the dynamic variation in noise.

Two figures of merits (FoMs) determine the quality of each mesh topology with respect to their tolerance to phase uncertainty and loss. To obtain these FoMs, an ONN is first trained, then made to classify the validation dataset under varying experimental conditions. The FoMs define a surface area of the ONN simulation results that achieves 75% classification accuracy achieved. If phase uncertainty is taken into account (i.e., all that is considered as $\sigma_\theta$ and $\sigma_\phi$), the FoM is in units of radian squared (rad²). On the other hand, if the parameters varied include both loss of the constituent MZIs and the phase uncertainty $\sigma_\theta$ and $\sigma_\phi$, the FoM is in units of decibel multiplied by radian (dB.rad). In this situation, we assume $\sigma_\theta$ is equal to $\sigma_\phi$. The two presented FoMs enables the study of ONNs' ability to handle practical uncertainties. This analysis has stochastic components (the loss varies depending on fabrication quality and position of the MZI on the chip, for example). Thus, every sample in the validation dataset is retested multiple times, and the average classification accuracy is taken.

Figure 5 (a-d) demonstrates the ability of the $10 \times 10$ meshes to handle the $\theta$ and $\varphi$ phase uncertainties. This figure presents the simulation results for classification of the Gaussian dataset when the MZI loss is 0 dB. Details of simulation parameters are presented in the supplementary materials. In all meshes, the accuracy degrades with the phase uncertainty, however, Clements and Bokun are more robust to phase error. The shorter optical depth in Clements and Bokun mainly contributes to the improved robustness to the phase uncertainties. Figure 5 (e-h) shows the meshes phase uncertainty tolerance with $\sigma_\theta = \sigma_\phi$ and resistance to loss of the constituent MZIs. Clements and Bokun with shorter optical depth demonstrate better loss tolerance compared to the two other counterparts. Diamond with symmetrical structure shows relatively better performance in terms of loss tolerance compared to the Reck.

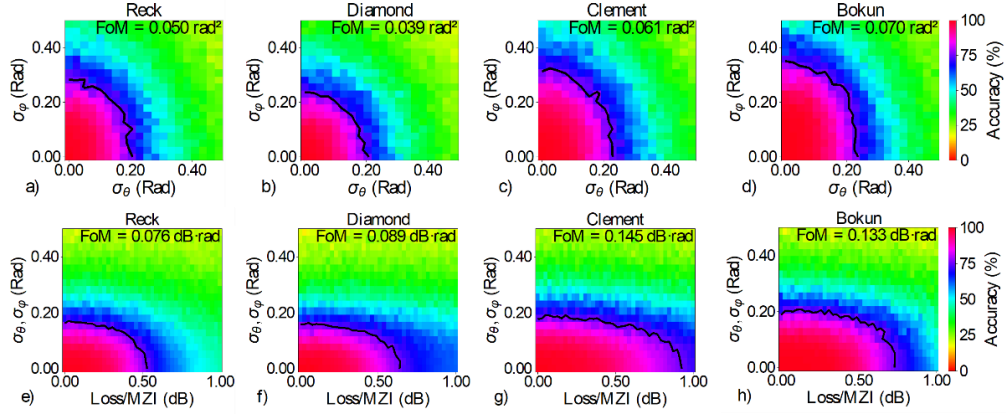

Fig. 5. Classification accuracy of a $10 \times 10$ ONNs for the Gaussian dataset. (a)—(d) are based on the Reck, Diamond, Clements, and Bokun mesh topologies with varying $\theta$ and $\varphi$ phase uncertainty for 0 dB loss per MZI. (d)—(f) shows the classification accuracy for $\sigma_\theta = \sigma_\phi$ and resistance to MZIs loss. The contour (black line) shows the FoM representing the area of above 75% classification accuracy.

Figure 6 (a-d) presents the classification accuracy of a two-layered $10 \times 10$ MNIST classifier based on the four presented architectures in the presence of $\theta$ and $\varphi$ phase errors. In the case of the MNIST classification, since the classification is more complex, we used a two-layered network to increase the classification accuracy. Simulation results of a single layer $10 \times 10$ classifier is also provided in the supplementary materials. Similar to the case of the Gaussian dataset, Clements and Bokun are more robust to phase error. Figure 6 (e-h) shows the classification accuracy in the presence of phase error and loss. The Clements and Bokun with minimum optical depth provide more robustness against optical loss, i.e., 0.081 dB.rad and 0.049 dB.rad, respectively. However, the classification accuracy of the Reck and Diamond decreases with insertion loss leading to a reduction in the FOM, i.e., 0.018 dB.rad and 0.019 dB.rad, respectively.

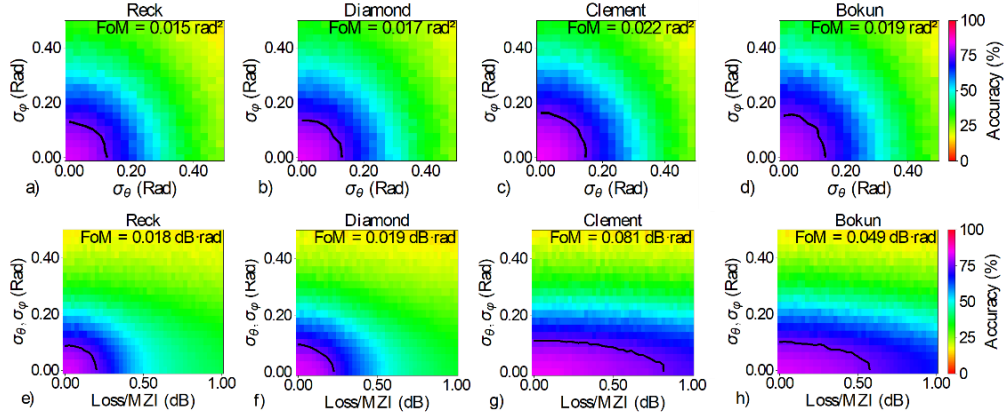

Fig. 6. Classification accuracy of a two-layered 10 × 10 ONNs for the MNIST dataset. (a)—(d) are based on the Reck, Diamond, Clements, and Bokun mesh topologies with varying $\theta$ and $\varphi$ phase uncertainty for 0 dB loss per MZI. (d)—(f) shows the classification accuracy for $\sigma_\theta = \sigma_\phi$ and resistance to MZIs loss. The contour shows the FoM representing the area of above 75% classification accuracy.

To better see the relation between classification accuracy and phase setting of phase shifters we may translate the phase error of phase shifters to the temperature error through the following equation [34]:

$$\Delta\Theta = \frac{2\pi L}{\lambda_0} \frac{dn}{dT} \Delta T \qquad (12)$$

in which $\Delta\Theta$ is the phase error (for $\theta$ or $\varphi$ phase shifter), $L$ is the phase shifter length, $\lambda_0$ is the wavelength, $dn/dT$ is the thermo-optic coefficient of the phase shifter, and $\Delta T$ is the temperature variation error. Through eq. 12, we see that in a typical 100 µm long phase shifter with a thermo-optic coefficient of $1.8 \times 10^{-4}$ K$^{-1}$, a 2.7 K variation of temperature leads to 0.2 rad change in the phase shift at 1550 nm wavelength. The similar phase shifter requires approximately 43 Kelvin variation in temperature for $\pi$ phase shift. Due to the thermal crosstalk between phase shifters, maintaining the temperature of a phase shifter in the range of ±2 K while the temperature of the adjacent phase shifter (with a few microns proximity) may vary up to 43 K is challenging. Moreover, fabrication variations such as waveguide sidewall roughness also adds to the phase error.

## 5. Discussion on Energy Efficiency

The energy consumption (in the unit of Joule per operation) for an $N \times N$ mesh of interferometers can be calculated as [36]:

$$E_{static}(J/Op) = \frac{n \times P_{PS}}{N^2 \, VR} \qquad (13)$$

In which $n$ is the total number of phase shifters in the mesh, $P_{PS}$ is the power dissipation in a phase shifter, and VR is the rate of the input vector being multiplied. We consider VR of 10 *GOp/s*. Although the interferometer-based optical processor performs analog vector matrix multiplication, the input vector is mainly changing with a specific rate, hence the computation speed is limited by the speed of incoming vectors. Figure 7 (a) compares the energy consumption of four the 10 × 10 different optical processor structures: the Reck, Diamond, Clements, and Bokun. We considered a power dissipation of 20 mW/$\pi$ for the thermos-optic phase shifters (TOPS) on silicon on insulator (SOI) platform [25, 35]. Assuming uniform distribution of phases, the average power consumption of a TOPS is $P_\pi/2$ where $P_\pi$ is the power required for the $\pi$ phase shift. As shown in Figure 7 (a), Reck and Clements structures with a smaller number of MZIs show better efficiency for static weight matrix while Diamond and Bokun dissipate more energy in larger number of MZIs.

The energy consumption presented by eq. 13 does not include the energy dissipated during the programming phase. This equation only represents the case of using static weight matrix, when the time/energy required for programming compared to that for computation is negligible. If the weight matrix changes more often, we must, however, include the energy dissipation of the programming phase. Let us assume the weight matrix changes with the frequency of $f_w$, and we spend the time of $t_{Prog}$ for programming. The modified energy consumption considering the energy spent for the programming becomes:

$$E_{total} = \frac{\frac{1}{f_w} - t_{prog}}{\frac{1}{f_w}} E_{static} = (1 - f_w \cdot t_{prog}) E_{static} \tag{14}$$

Employing faster programming methods as well as increasing phase shifter speed reduce $t_{Prog}$ and reduce the total energy consumption. The in-situ programming relying on optimization techniques performed on the chip increases the $t_{Prog}$ hence $E_{total}$. Figure 7 (b) compares $E_{total}$ for the four meshes. For Reck and Clements, we assumed a backpropagation programming method with 200 iterations [16]. The ex-situ programming (i.e., predefined weight matrix programming) is not viable on these meshes due to the lack of monitoring options. For Diamond and Bokun, we considered ex-situ programming with 10 iterations/MZI monitoring the MZI state and readjusting their bias. The programming time is estimated based on a 2.2 µs transit time in TOPS [35]. The theoretical optical training time of each iteration needs to account for the maximum of the TOPS transit time and the electronic delay [37]. Considering the slow response of TOPS, we can neglect the electronic delays. The power consumption required by the electronics is not accounted in eq. (14). We assumed the electronic power consumption is negligible compared to the power-hungry TOPS. More accurate estimation of $E_{total}$ may include the electronic energy consumption by simply adding this value to the $E_{static}$. Adding the electronic power consumption will change the scale of Fig. 7 (b) while the trend will remain almost the same. From Fig. 7, for Clements the energy consumption increases from 450 fJ/Op for stationary weight matrix to 3750 fJ/Op for weight matrix changing at 2 kHz. While for Bokun, taking the advantage of monitoring feature provided by the architecture, the energy consumption slightly increases from 610 fJ/Op to 638 fJ/Op. Diamond mesh also saves energy in the training, however, as shown in Figs 5–6, its longer optical depth deteriorates its performance in presence of optical insertion loss and phase error.

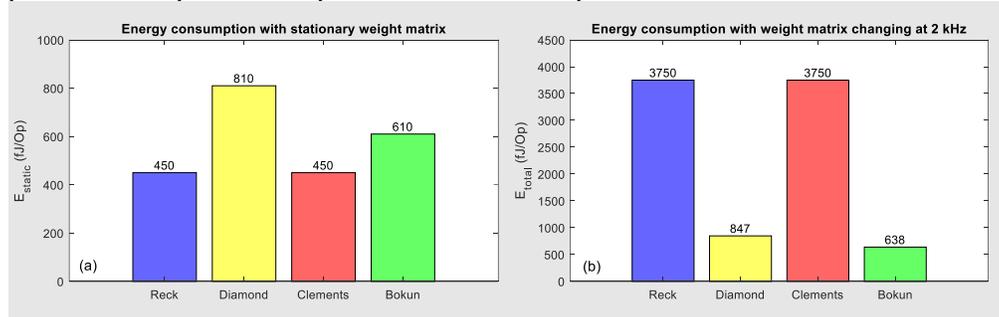

Fig. 7. Energy consumption in units of energy per operation (a) with and (b) without programming.

## 6. Summary

Bokun mesh, proposed in this work, is a topology arrangement that merges the attributes of the prior processor topologies Diamond and Clements for optical processors. Like Diamond, Bokun provides diagonal path going through every individual MZI enabling phase monitoring. Providing the monitoring option, Bokun's programming is faster improving the total energy efficiency of the processor. Unlike Diamond, Bokun maintains the minimum optical depth making it more resilient to the loss and fabrication process imperfections.

**Acknowledgment**

The authors thank Simon Geoffroy-Gagnon for configuring Neuroptica in 2021 in the first phase of this project.